\documentclass[aps,prl,twocolumn,groupedaddress,showpacs,superscriptaddress,nofootinbib]{revtex4-1}
\usepackage{graphicx}
\usepackage{color}
\usepackage{xurl}
\usepackage{hyperref}
\usepackage{epigraph}
\begin{document}
\title{Is Gravity a Real Force or Merely a Geometrical Effect?}
\epigraph{
  Denying relativity in the age of precision experiments is like watching a bird take off and still claiming that flight is impossible.
} {V. G. Rousseau}

\author{V.G.~Rousseau}
\affiliation{Physics Department, Xavier University of Louisiana, 1 Drexel Dr., New Orleans, LA 70125, USA}

\relpenalty=10000      
\binoppenalty=10000

\begin{abstract}
Whether gravitation should be regarded as a fundamental force or as a manifestation of spacetime geometry remains an interesting conceptual question in physics. In Newtonian mechanics, gravitation is described as a force acting between massive objects, whereas General Relativity describes gravitational phenomena in terms of spacetime curvature, with freely falling bodies following geodesics. With this fundamental difference, it is legitimate to ask whether gravity is a real force or not, and it is very common to receive from experts a somewhat condescending answer such as \textit{``Drop a rock on your foot and you will feel that the force is real"}. This paper does not attempt to determine whether gravity is fundamentally a force or a geometrical phenomenon. Rather, it questions whether commonly invoked everyday experiences provide valid grounds for deciding between these interpretations.
\end{abstract}

\maketitle
\section{Introduction}
Few concepts in physics are as intuitively familiar as gravity. Objects fall toward Earth, planets orbit the Sun, and we experience weight whenever we stand on the ground. In Newtonian mechanics, these phenomena are explained through gravitational forces acting between massive bodies. This description has been extraordinarily successful and remains widely used in physics and engineering.

However, Einstein's General Theory of Relativity (GR) introduced a fundamentally different interpretation\cite{Einstein1,Einstein2,Misner,Wald,Will}. Rather than describing gravitation as a conventional force acting between objects, GR describes gravitational phenomena through the curvature of spacetime. In this framework, freely falling objects follow geodesics, which represent their natural trajectories through curved spacetime. An object in free fall therefore experiences no proper acceleration, even though its position relative to the Earth changes continuously.

This difference between the Newtonian and relativistic descriptions raises a fundamental question: Should gravitation be regarded as a real force, or is it more appropriately understood as a manifestation of spacetime geometry? When asking if gravity is a real force, it is not uncommon to receive from experts a response that involves a form of \textit{proof by condescension}. For example, Viktor T. Toth, who is widely recognized as an expert in General Relativity and gravitation, says\cite{Toth}:
\begin{quote}
  \textit{So don’t let anyone tell you that gravity ``is not a force at all". If they try, just tell them to drop half a brick (a light one, we don’t want them to suffer permanent injury) on their big toe and then let them explain to you how they felt no force at all. I doubt they’ll succeed.}
\end{quote}

The response strongly suggests that the reality of gravity as a force should be self-evident. However, even though Einstein cautioned us not to take the geometric description of gravitation too literally, both the motion of the rock and the pain induced on impact or when the rock is resting on the foot are perfectly well explained in that framework without involving a gravitational force at all, as we shall show. By considering both a falling rock and a rock resting on someone's foot, we demonstrate that the experience of weight does not, by itself, establish gravitation as a fundamental force.

The objective is not to establish that one interpretation is necessarily correct, but rather to question whether the arguments commonly invoked in support of the force interpretation are physically conclusive.

\section{The Force Experienced When a Rock Falls on the Foot}
Consider a rock of mass $m$ falling from a certain height toward a person's foot. Upon impact, the person experiences a force that can be quite substantial, potentially causing injury. At first glance, this seems to provide an intuitive demonstration that gravity is a real force. After all, without gravity, the rock would not have fallen onto the foot. However, this reasoning overlooks the physical origin of the force actually experienced by the person.

Let us examine the collision more carefully. Immediately before impact, the rock is moving downward. During the collision, it undergoes a rapid deceleration as it is brought to rest by the resistance of the foot. This situation satisfies the conditions of the Newtonian limit (non-relativistic speed, weak gravitational field, static field), so the familiar form of Newton's second law remains an excellent approximation,
\begin{equation}
  \vec{F}_{\mathrm{net}}=m\vec{a},
\end{equation}
where $\vec{F}_{\mathrm{net}}$ is the net force acting on the rock.

The dominant force responsible for this deceleration is the upward contact force exerted by the foot on the rock. Likewise, the rock exerts a similar contact force on the foot (with equal magnitude and opposite direction in the approximation of Newtonian mechanics).

What is the physical origin of this contact force? At the microscopic level, it arises predominantly from electromagnetic interactions between the atoms and electrons constituting the rock and those constituting the foot. The resistance to compression involves electromagnetic interactions and quantum-mechanical effects associated with the electronic structure of matter. Thus, the force experienced by the foot during the collision is predominantly electromagnetic in origin. It is not the gravitational attraction between the Earth and the rock.

Of course, gravity is responsible for the rock acquiring its downward velocity in the Newtonian description. In General Relativity, the rock acquires this velocity while following a geodesic through curved spacetime. However, neither description requires the force experienced during the collision to be gravitational in origin.

The distinction is important: The mechanism responsible for bringing the rock toward the foot is not the same as the mechanism responsible for the force experienced when the rock strikes it. Consequently, the familiar rock-on-the-foot argument does not establish that gravity is a fundamental force. It establishes only that a falling object can produce a substantial contact force of electromagnetic origin upon colliding with another object.

One might object that the preceding example concerns a collision and that the situation is different when a rock simply rests on someone's foot. Indeed, a stationary rock exerts a continuous force on the foot. This is conventionally interpreted as the rock's weight, and it might appear to provide a more convincing argument for the existence of gravitational force.

In the following sections, we analyze the situation rigorously in the framework of Newtonian mechanics, then in the framework of General Relativity.

\section{Newtonian Description}
\subsection{The acceleration of the free-falling rock}
\begin{figure}[h]
  \centerline{\includegraphics[width=0.5\textwidth]{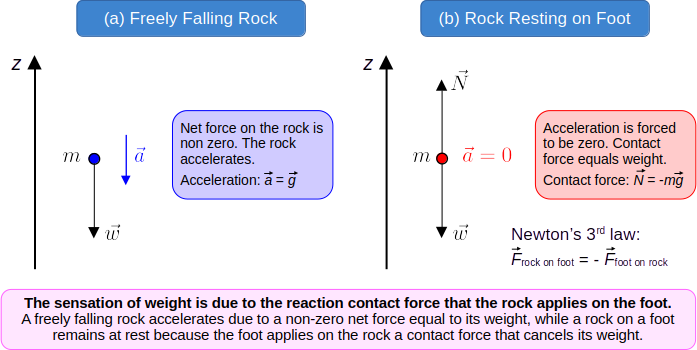}}
  \caption{Newtonian description using free-body diagrams of the rock when it is (a) freely falling and (b) resting on a supporting foot. When it is free-falling, the rock experiences only its own weight, therefore it accelerates downward according to Newton's second law. When it is resting on the foot, its acceleration is forced to be zero by the upward contact force that the foot exerts on it, which cancels its weight. According to Newton's third law, the rock applies a reaction contact force equal in magnitude and opposite in direction, thus downward, on the foot.}
  \label{FreeBodyDiagrams}
\end{figure}

In the Newtonian description of a free-falling rock, there is a downward gravitational force applied to the rock (its weight $\vec w$), hence a net force. This is illustrated by the free-body diagram of Fig.~\ref{FreeBodyDiagrams} (a). According to Newton's second law, this net force results in a downward acceleration:
\begin{equation}
  \label{NewtonianAcceleration} \vec{F}_{\mathrm{net}}=\underbrace{\vec w}_{m\vec g}=m\vec a\quad\Rightarrow\quad \vec a=\vec g
\end{equation}
In this view, a gravitational force is clearly responsible for the rock's motion.

\subsection{Rock resting on the foot}
When the rock is resting on the foot, the gravitational force on the rock does not disappear. But now the rock's acceleration is forced to be zero. In order to preserve Newton's second law, we must conclude that the foot applies on the rock an upward force $\vec N$ that cancels its weight, Fig.~\ref{FreeBodyDiagrams} (b). This force would not exist if the foot wasn't in contact with the rock. Thus, we have just established that the force that the foot exerts on the rock, commonly known as the \textit{normal force}, is a contact force:
\begin{equation}
  \vec{F}_{\mathrm{net}}=\vec N+\vec w=\underbrace{m\vec a}_0\quad\Rightarrow\quad \vec N=-m\vec g
\end{equation}
If the foot exerts an upward contact force on the rock, then Newton's third law states that the rock must exert a reaction force that is opposite in direction and equal in magnitude, hence downward:
\begin{equation}
  \vec{F}_\textrm{\tiny rock on foot}=-\vec{F}_\textrm{\tiny foot on rock}=m\vec g
\end{equation}
Note that the fact that the force exerted by the rock on the foot is equal to the rock's weight does not establish that it is gravitational. Rather, it is a contact force, therefore of electromagnetic nature. However, this force would not be present in the absence of a gravitational force. So, in this Newtonian description, it seems legitimate to conclude that the pain induced in the foot is ultimately due to the gravitational force on the rock.

\section{General Relativistic Description}
The situation becomes particularly interesting when examined within the framework of General Relativity. In this theory, gravitational phenomena are described through the geometry of spacetime. Freely falling bodies follow geodesics, and their motion does not require the introduction of a conventional gravitational force.

Consider the four-position $x^\mu$ of the rock, where $x^0=ct$ is the time component and $x^1,x^2,x^3$ are the spatial components. The rock's four-velocity is the tangent vector,
\begin{equation}
  \label{Four-velocity} u^\mu=\frac{dx^\mu}{d\tau},
\end{equation}
where $\tau$ is the rock's proper time. Its four-acceleration is given by the covariant derivative of the four-velocity,
\begin{equation}
  \label{Four-acceleration} a^\mu=\frac{du^\mu}{d\tau}+\Gamma_{\alpha\beta}^\mu u^\alpha u^\beta,
\end{equation}
where $\Gamma_{\alpha\beta}^\mu$ are the Christoffel symbols given in terms of the metric tensor $g_{\mu\nu}$ and its inverse $g^{\mu\nu}$ as:
\begin{equation}
  \Gamma_{\alpha\beta}^\mu=\frac12 g^{\lambda\mu}\Big(\frac{\partial g_{\lambda\beta}}{\partial x^\alpha}+\frac{\partial g_{\lambda\alpha}}{\partial x^\beta}-\frac{\partial g_{\alpha\beta}}{\partial x^\lambda}\Big)
\end{equation}
The proper acceleration is a general invariant defined as the magnitude of the four-acceleration. Using the mostly plus signature $(-,+,+,+)$, it is given by:
\begin{equation}
  \label{ProperAcceleration} \alpha=\sqrt{g_{\mu\nu}a^\mu a^\nu}
\end{equation}
For the description of the rock in the gravitational field of the Earth, we can approximate the Earth as a perfect non-rotating sphere of radius $R$ and mass $M$. With this model, and using spherical coordinates $r,\theta,\varphi$, the Schwarzschild metric is the exact solution to Einstein's field equations for radial distances greater or equal to the radius $R$. The metric tensor is given by
\begin{equation}
  g_{\mu\nu}=\left(\begin{array}{cccc}-\Big(1-\frac{r_s}{r}\Big)&0&0&0\\0&\Big(1-\frac{r_s}{r}\Big)^{-1}&0&0\\0&0&r^2&0\\0&0&0&r^2\sin^2\!\theta\end{array}\right),
\end{equation}
where $r_s$ is the Schwarzschild radius,
\begin{equation}
  \label{SchwarzschildRadius} r_s=\frac{2GM}{c^2},
\end{equation}
and $G$ is the universal gravitational constant. For this metric, only 9 unique Christoffel symbols out of 64 are nonzero (excluding symmetric swaps like $\Gamma_{\alpha\beta}^\mu=\Gamma_{\beta\alpha}^\mu$):
\begin{eqnarray}
  && \Gamma_{tr}^t=\Gamma_{rt}^t=\frac{r_s}{2r(r-r_s)}\\
  && \Gamma_{tt}^r=\frac{r_s}{2r^2}\Big(1-\frac{r_s}{r}\Big)\\
  && \Gamma_{rr}^r=-\frac{r_s}{2r(r-r_s)}\\
  && \Gamma_{\theta\theta}^r=-r\Big(1-\frac{r_s}{r}\Big)\\
  && \Gamma_{\varphi\varphi}^r=-r\sin^2\!\theta\Big(1-\frac{r_s}{r}\Big)\\
  && \Gamma_{r\theta}^\theta=\Gamma_{\theta r}^\theta=\frac{1}{r}\\
  && \Gamma_{\varphi\varphi}^\theta=-\sin\theta\cos\theta\\
  && \Gamma_{r\varphi}^\varphi=\Gamma_{\varphi r}^\varphi=\frac{1}{r}\\
  && \Gamma_{\theta\varphi}^\varphi=\Gamma_{\varphi\theta}^\varphi=\cot\theta
\end{eqnarray}
The spacetime interval $ds^2=g_{\mu\nu}dx^\mu dx^\nu$ takes the form:
\begin{eqnarray}
  \nonumber             ds^2 &=& -\Big(1-\frac{r_s}{r}\Big)c^2dt^2+\Big(1-\frac{r_s}{r}\Big)^{-1}dr^2\\
  \label{SpacetimeInterval}      &&  +r^2d\theta^2+r^2\sin^2\!\theta\:d\varphi^2
\end{eqnarray}
By definition of the rock's infinitesimal proper time $d\tau$, we must have $ds^2=-c^2d\tau^2$. It follows that the generalized time dilation factor $\mathcal D=\frac{dt}{d\tau}$ takes the form
\begin{equation}
  \label{TimeDilationFactor} \mathcal D=\frac{1}{\sqrt{1-\frac{r_s}{r}-\frac{u_r^2}{c^2\Big(1-\frac{r_s}{r}\Big)}-\frac{r^2u_\theta^2}{c^2}-\frac{r^2\sin^2\!\theta u_\varphi^2}{c^2}}},
\end{equation}
where $u_r=\frac{dr}{dt}$, $u_\theta=\frac{d\theta}{dt}$, and $u_\varphi=\frac{d\varphi}{dt}$.

\subsection*{Free-falling rock}
If the rock is in a free fall, then it follows by definition a geodesic described by the equation:
\begin{equation}
  \label{Geodesic} \frac{d^2x^\mu}{d\tau^2}+\Gamma_{\alpha\beta}^\mu\frac{dx^\alpha}{d\tau}\frac{dx^\beta}{d\tau}=0
\end{equation}
Substituting (\ref{Four-velocity}) in (\ref{Geodesic}) and identifying the resulting left-hand side as (\ref{Four-acceleration}), we see that the free-falling rock's four-acceleration is zero and, consequently, has a zero proper acceleration:
\begin{equation}
  a^\mu=0 \quad\Rightarrow\quad \alpha=0
\end{equation}
In the rock's instantaneous local inertial frame, Newton's second law, $\vec F_\mathrm{net}=\frac{d\vec p}{dt}$, reduces to its classical form, $\vec F_\mathrm{net}=m\vec a$. Since the acceleration in this frame is equal to the proper acceleration, we conclude that the net force on the rock is zero:
\begin{equation}
  \vec F_\mathrm{net}=0\qquad \textrm{No acting force}
\end{equation}
This is consistent with the equivalence principle underlying General Relativity: A freely falling observer experiences no gravitational force locally, as confirmed by an accelerometer reading zero.

It is instructive to determine how an observer at rest with respect to the ground and at low altitude, $r=R$, perceives the rock's acceleration at the instant it is released from the same altitude. For this, we consider the radial component of the geodesic equation (\ref{Geodesic}):
\begin{equation}
  \frac{d^2r}{d\tau^2}+\Gamma_{tt}^r\bigg(\frac{d(ct)}{d\tau}\bigg)^2+\Gamma_{rr}^r\bigg(\frac{dr}{d\tau}\bigg)^2=0
\end{equation}
At the instant of release, $u_r=u_\theta=u_\varphi=0$ and $r=R$, which implies $\frac{dr}{d\tau}=0$ and $\frac{dt}{d\tau}=1/\sqrt{1-\frac{r_s}{R}}$. Substituting in the above equation, we obtain:
\begin{equation}
  \label {Math-1} \frac{d^2r}{d\tau^2}_{|\textrm{\tiny release}}=-\frac{r_sc^2}{2R^2}
\end{equation}
The observer measures acceleration using proper length $dl$ and proper time $dT$:
\begin{equation}
  \label{ObserverAcceleration} a_\textrm{\tiny obs}=\frac{d^2l}{dT^2}_{|\textrm{\tiny release}}
\end{equation}
He remains at $r=R$, so his proper time is obtained from (\ref{TimeDilationFactor}) as:
\begin{equation}
  dT=\sqrt{1-\frac{r_s}{R}}dt
\end{equation}
Meanwhile, the rock's radial proper distance $dl$ from the release point is locally obtained by substituting $dt=d\theta=d\varphi=0$ in the spacetime interval (\ref{SpacetimeInterval}):
\begin{equation}
  \label{Math-2} dl=\frac{dr}{\sqrt{1-\frac{r_s}{r}}}
\end{equation}
By using the chain rule and the product rule, the second derivative $\frac{d^2l}{dT^2}$ in (\ref{ObserverAcceleration}) can be expressed in terms of $\frac{d^2r}{d\tau^2}$. The expression simplifies greatly when it is evaluated at the instant of release for which $dT=d\tau$, $r=R$, and $\frac{dr}{d\tau}=0$. This leads to:
\begin{equation}
  a_\textrm{\tiny obs}=\frac{1}{\sqrt{1-\frac{r_s}{R}}}\frac{d^2r}{d\tau^2}_{|\textrm{\tiny release}}
\end{equation}
Substituting (\ref{Math-1}) and the expression for the Schwarzschild radius (\ref{SchwarzschildRadius}), we conclude that:
\begin{equation}
  \label{Math-3} a_\textrm{\tiny obs}=-\frac{GM}{R^2\sqrt{1-\frac{2GM}{Rc^2}}}\approx -9.82\:{\rm m/s^2}
\end{equation}
As a result, we recover the Newtonian result (\ref{NewtonianAcceleration}) up to a small relativistic correction: Close to the ground, an observer at rest measures a downward acceleration with a magnitude nearly equal to $g=\frac{GM}{R^2}$.

\subsection{Rock resting on the foot}
If the rock is resting on the foot, then its spatial velocity components are zero, the radial position is constant, $r=R$, and Eq.~(\ref{TimeDilationFactor}) reduces to:
\begin{equation}
  \mathcal D=\frac{1}{\sqrt{1-\frac{r_s}R}}
\end{equation}
The rock's four-velocity $u^\mu=\frac{dx^\mu}{d\tau}=\mathcal D\frac{dx^\mu}{dt}$ is constant,
\begin{equation}
  u^\mu=\bigg(\frac{c}{\sqrt{1-\frac{r_s}R}},0,0,0\bigg),
\end{equation}
and its derivative with respect to proper time is zero, $\frac{du^\mu}{d\tau}=0$. As a result, only the radial component of the four-acceleration (\ref{Four-acceleration}) is nonzero, $a^\mu=\Gamma_{tt}^r(u^t)^2$, leading to:
\begin{equation}
  a^\mu=\Big(0,\frac{r_sc^2}{2R^2},0,0\Big)
\end{equation}
The rock's proper acceleration, $\alpha=\sqrt{g_{\mu\nu}a^\mu a^\nu}=\sqrt{g_{rr}}a^r$, is therefore:
\begin{equation}
  \alpha=\frac{r_sc^2}{2R^2\sqrt{1-\frac{r_s}{R}}}=\frac{GM}{R^2\sqrt{1-\frac{2GM}{Rc^2}}}\approx 9.82\:{\rm m/s^2}
\end{equation}
As a result, an accelerometer attached to the rock reads an upward acceleration with a magnitude exactly equal to that of the free-falling rock (\ref{Math-3}). Here again, in the rock's instantaneous local inertial frame, Newton's second law holds true in its classical form. We conclude that the foot exerts on the rock an upward contact force:
\begin{equation}
  \vec F_\mathrm{net}=\frac{GMm}{R^2\sqrt{1-\frac{2GM}{Rc^2}}}\hat r\qquad \textrm{Contact force}
\end{equation}
The upward contact force exerted by the foot on the rock is accompanied by a downward contact force exerted by the rock on the foot. These forces have equal magnitudes to an excellent approximation. The downward force produces mechanical stress in the tissues of the foot and is responsible for the sensation of weight. The General Relativity interpretation provides a particularly interesting perspective on this situation: The rock is not being held in place against gravity, it is being prevented by the foot from following its natural geodesic trajectory through spacetime.

\begin{figure}[h]
  \centerline{\includegraphics[width=0.5\textwidth]{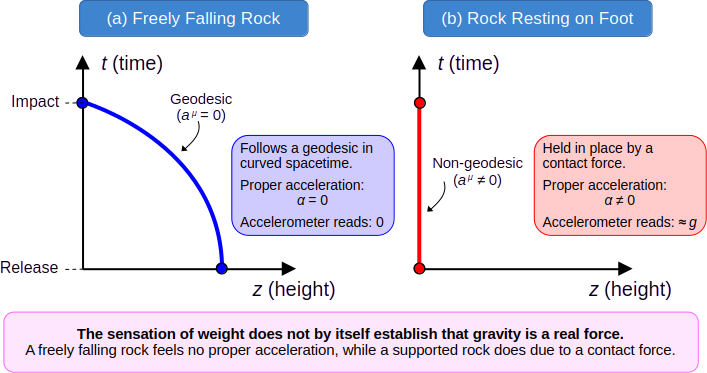}}
  \caption{Spacetime diagrams illustrating the trajectories of (a) a freely falling rock and (b) a rock resting on a supporting foot. Time increases upward, and the horizontal coordinate represents the rock's height above the foot. The freely-falling rock follows a geodesic and has zero proper acceleration, whereas the supported rock follows a non-geodesic worldline and has nonzero proper acceleration due to the contact force.}
  \label{SpacetimeDiagrams}
\end{figure}

The distinction between these two situations is illustrated in Fig.~\ref{SpacetimeDiagrams}, which compares the worldlines of a freely falling rock and a rock resting on a supporting surface in Earth-fixed coordinates. Although the freely falling rock experiences coordinate acceleration, its proper acceleration is zero because it follows a geodesic. Conversely, the rock resting on the foot has a nonzero proper acceleration because the contact force prevents it from following its natural free-fall trajectory.

\section{Back to the Equivalence Principle}
The preceding examples are closely related to the equivalence principle, one of the fundamental ideas underlying General Relativity.

\begin{figure}[h]
  \centerline{\includegraphics[width=0.5\textwidth]{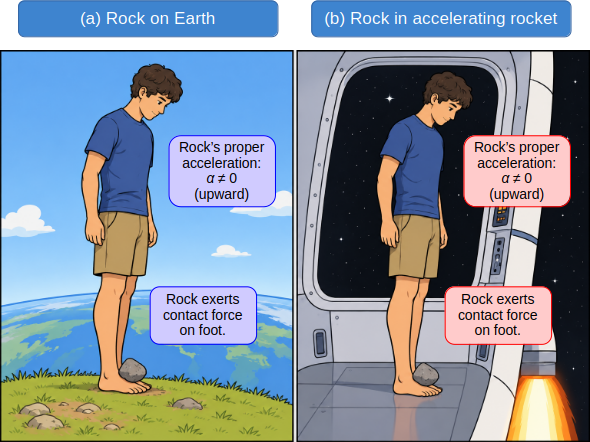}}
  \caption{The equivalence principle - Locally, the rock resting on the foot of the observer experiences a proper acceleration, whether the observer and the rock are on Earth or in an accelerating rocket away from any gravitational fields. In both cases, the rock exerts a contact force on the foot. The sensation felt by the foot therefore cannot constitute conclusive evidence that gravity is a force.}
  \label{EquivalencePrinciple}
\end{figure}

Consider the two observers in Fig.~\ref{EquivalencePrinciple}:
\begin{itemize}
  \item An observer standing on Earth's surface with a rock resting on their foot (a).
  \item An observer inside a rocket accelerating upward in otherwise empty space, far from any gravitational source, with a rock on their foot (b).
\end{itemize}
Locally, the physical phenomena experienced by these observers are essentially identical. This is what Einstein famously called \textit{``the happiest thought of my life"}. Both feel pressed against their supporting surfaces, and both experience an object resting on their feet as having weight. In the first situation, the observer and the rock are stationary relative to Earth. In the second, the observer and the rock are accelerating relative to an inertial frame in the absence of a gravitational field. The equivalence principle states that these situations are locally indistinguishable through any experiments. This provides another reason why the sensation of weight cannot, by itself, establish the existence of a gravitational force. If both observers experience the same physical sensation from the rock on their foot, and yet one of them occurs in a situation where no gravitational field is present, then that sensation cannot constitute conclusive evidence that gravity is a force.

\section{Conclusion}
The preceding analysis suggests that the everyday association between gravity and the sensation of weight is conceptually misleading. It is entirely possible to describe gravitational phenomena through a mathematical formalism in which gravity does not act as a conventional force, regardless of whether the object under consideration is in a free fall or not. This does not imply that gravity is fictitious in the same sense as centrifugal or Coriolis forces. Spacetime curvature has measurable physical consequences that cannot be eliminated through a simple change of reference frame. For example, tidal effects arise from spacetime curvature and can be described through geodesic deviation. These effects distinguish a genuine gravitational field with nonzero curvature from a uniformly accelerating reference frame. Thus, rejecting the rock-on-the-foot argument does not amount to denying the physical reality of gravitational phenomena. It simply means that the existence of these phenomena does not, by itself, determine whether gravity should be interpreted as a fundamental force or as a geometrical property of spacetime.

It is important, however, to recognize two facts that cannot be ignored:
\begin{itemize}
  \item Precision experiments confirm that the purely geometrical description of gravity provided by General Relativity supersedes the force description from Newtonian mechanics.
  \item All attempts that have been made for unifying the expected fundamental gravitational force with the three other fundamental forces have been unsuccessful so far, and no graviton has ever been observed. Of course, this doesn't mean that it won't eventually happen.
\end{itemize}
Given these facts, it is not unreasonable to consider the possibility that the geometrical description should actually be taken literally. In the search for a unifying theory, instead of trying to quantize General Relativity, perhaps we should try to gravitationalize Quantum Mechanics!

\end{document}